\documentclass[aps,prb,10pt,twocolumn,superscriptaddress,floatfix,longbibliography]{revtex4-2}
\usepackage{comment}
\usepackage{amsmath,amssymb,bm}
\usepackage{graphicx}

\graphicspath{{./}}   
\usepackage[colorlinks=true,allcolors=blue]{hyperref}

\newcommand{\Dnl}{D_{\mathrm{NL}}}          
\newcommand{\tauM}{\tau^{M}}                          
\newcommand{\DM}{D^{M}_{\mathrm{NL}}}         
\newcommand{\LA}{L_A}
\newcommand{\tr}{\mathrm{tr}}
\newcommand{\MF}{\mathcal{M}_{\mathrm{FNL}}}       
\usepackage{orcidlink}

\usepackage{xcolor}
\newif\ifshowrevisions
\showrevisionstrue
\DeclareRobustCommand{\rev}[1]{%
  \ifshowrevisions{\color{red}\hypersetup{allcolors=red}#1}\else#1\fi}
\newenvironment{revision}{%
  \begingroup\ifshowrevisions\color{red}\hypersetup{allcolors=red}\fi
}{\endgroup}
\begin{document}

\title{Nonlocal Magic Spreading in Many-body Quantum Dynamics:\\
From Chaotic Evolution to Quasi-particle Picture in Integrable Models}

\author{Sreemayee Aditya~\orcidlink{0000-0002-0412-7944}}
\affiliation{Institut f\"ur Theoretische Physik, Universit\"at zu K\"oln,
Z\"ulpicher Stra\ss e 77, 50937 K\"oln, Germany}
\author{Piotr Sierant~\orcidlink{0000-0001-9219-7274}}
\affiliation{Barcelona Supercomputing Center, Barcelona 08034, Spain}
\author{Xhek Turkeshi~\orcidlink{0000-0003-1093-3771}}
\affiliation{Institut f\"ur Theoretische Physik, Universit\"at zu K\"oln,
Z\"ulpicher Stra\ss e 77, 50937 K\"oln, Germany}

\begin{abstract}
Entanglement and magic are resources that reveal complementary aspects of quantum many-body systems. Their interplay is captured by nonlocal magic, the magic that survives arbitrary local changes of basis. Yet their markedly different dynamical behavior leaves open how this irreducible component of magic spreads. Here we connect nonlocal magic to the capacity of entanglement, a tractable quantity measuring fluctuations of the entanglement Hamiltonian. This connection enables analytically controlled predictions across a wide range of many-body dynamics, from chaotic to integrable systems, which we investigate also using large-scale numerical simulations. In chaotic systems, nonlocal magic exhibits a transient buildup, with logarithmic growth in time followed by decay to a size-independent value. For integrable systems, we develop a quasiparticle picture in quantitative agreement with numerics, showing that the same initial growth instead leads to saturation at a value logarithmic in subsystem size. These contrasting behaviors have an operational consequence for entanglement embezzlement: the strongest scramblers embezzle only transiently, whereas free-fermionic dynamics can sustain universal embezzlement in the steady state.
\end{abstract}

\maketitle

\section{Introduction}

Entanglement~\cite{vedral2008entanglement,plenio2010colloquium} and magic~\cite{winter2022many,kitaev2005universal,emerson2014resource} are complementary quantum resources that underpin different approaches to the classical simulation of many-body systems: Tensor-network methods~\cite{schollwock2011density, Paeckel19, orus2019tensor, Cirac21mat} exploit limited entanglement, while stabilizer-based techniques exploit limited magic~\cite{howard2019simulation}. For example, product states admit trivial tensor-network representations yet can carry extensive magic, whereas Clifford circuits can generate extensive entanglement while remaining efficiently simulable~\cite{gottesman1998heisenberg,gosset2016improved,gottesman2004improved,gidney2021stim}.
This complementarity is also reflected in markedly different dynamical behavior. Entanglement typically grows linearly in time in both chaotic and integrable systems~\cite{Nahum17entanglement,nahum2020entanglement, turkeshi2023entanglement,calabrese2008evolution,cardy2005evolution}. Magic, by contrast, can globally equilibrate on much shorter timescales~\cite{sierant2025magic,sierant2025anticoncentration,turkeshi2025quantum,turkeshi2025clifford,denardis2025universality,turkeshi2025fermionic,sierant2024hilbert, garciasaez2026computing}, whereas locally it can exhibit transient growth followed by decay~\cite{Aditya2026growth,Aditya2026equivalence,turkeshi2026coherence}.

\begin{figure}[!htbp]
\includegraphics[width=0.9\columnwidth]{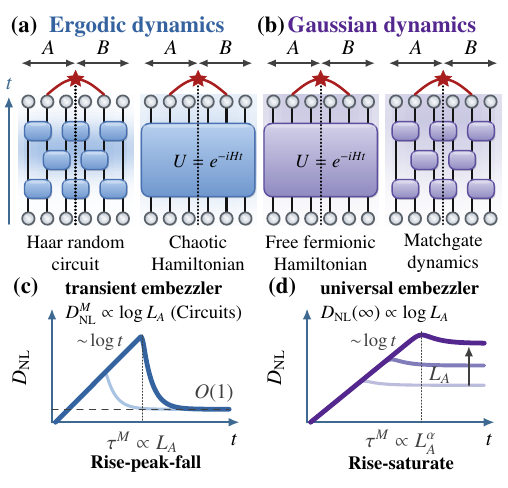}
\vspace{-0.4cm}
\caption{
\textbf{Many-body dynamics of nonlocal magic.}
Nonlocal magic $\Dnl$ 
for a subsystem of $L_A$ qubits of a 1D system under
(a)~ergodic dynamics---Haar random circuits and chaotic Hamiltonians---and
(b)~free-fermionic dynamics---free-fermionic Hamiltonians and matchgate circuits.
The blue shading fades as $\Dnl$ decays, whereas the persistent purple shading depicts its retention at late times.
(c)~Ergodic dynamics yields logarithmic growth of $\Dnl$ followed by a peak and decay to an $O(1)$ stationary value.
(d)~free-fermionic dynamics shares the logarithmic growth but retains a stationary value $\Dnl(\infty)\propto\log_2\LA$. Darker curves denote larger subsystems: their late-time values remain separated in (d), while they approach the same $O(1)$ baseline in (c). 
}
\label{fig:setup}
\end{figure}

Understanding their interplay requires identifying the magic that is irreducibly tied to entanglement. Nonlocal magic isolates this contribution by quantifying the magic that cannot be removed by local changes of basis on the two sides of a bipartition~\cite{wang2025quantum, oliviero2025gravitational, sierant2026exact, hamma2026local, sierant2026spectral}. This notion has found applications in holography~\cite{oliviero2025gravitational,cao2024trivial}, quantum fields and scattering~\cite{odavic2025harvesting,savage2026antiflatness}, and free-fermionic systems~\cite{turkeshi2026local,tirrito2026nonlocal}, with close connections to the entanglement spectrum~\cite{giampaolo2026local,giampaolo2026schmidt,cui2026entanglement,cui2026nonlocal,sierant2026spectral,wang2025quantum,liu2026entirely} and recent experimental measurements~\cite{hamma2025experimental}. However, evaluating nonlocal magic requires a demanding optimization over local bases, making this quantity difficult to explore in many-body systems. A recent measure based on stabilizer fidelity circumvents this obstacle through an exact solution of the optimization~\cite{sierant2026exact,hamma2026local}. The resulting quantity $\Dnl$ measures how closely the entanglement spectrum resembles that of a collection of Bell pairs. This advance opens the way to studying nonlocal magic dynamically, but the contrasting evolution of entanglement and magic leaves its fate unresolved: \textit{Does their irreducible combination follow either resource, or exhibit a distinct pattern of production and transport?} Answering this question remains challenging because $\Dnl$ depends on the full Schmidt spectrum, whereas theoretical descriptions of many-body dynamics typically provide access to only a few of its moments.

In this work, we make the dynamics of nonlocal magic tractable by relating $\Dnl$ to the capacity of entanglement $C$~\cite{Boer19capacity}, the second cumulant of the entanglement energy spectrum. This connection identifies spectral fluctuations as the physical ingredient controlling $\Dnl$ and enables predictions from a few cumulants, without reconstructing the full spectrum. The resulting description becomes exact in several scaling regimes, and we test its predictions against extensive numerical simulations of ergodic and free-fermionic dynamics, see Fig.~\ref{fig:setup}. Building on this connection, we develop a quasiparticle picture for nonlocal magic in integrable free-fermion systems, quantitatively capturing its production and transport after a quench. Both ergodic and free-fermionic dynamics exhibit logarithmic growth at early times, but their subsequent evolution differs sharply. In ergodic systems, $\Dnl$ reaches a transient maximum before decaying to an $O(1)$ stationary value at balanced bipartitions; in random circuits, this maximum scales as $\DM\propto\log_2\LA$ and occurs at $\tauM\propto\LA$. Free-fermionic systems instead retain nonlocal magic at late times, saturating at $\Dnl(\infty)\propto\log_2\LA$. The capacity explains this distinction: the ergodic steady states retain only $C=O(1)$, while the free-fermion steady states have an extensive capacity, $C\propto\LA$. Since diverging $D_{\mathrm{NL}}$ is equivalent to universal entanglement embezzlement~\cite{sierant2026exact}, these findings reveal that stronger scrambling can undermine entanglement embezzlement~\cite{hayden2003universal,gour2024complete, wilming2025critical}: by flattening the entanglement spectrum, the strongest scramblers retain universal embezzling power only transiently, whereas free-fermionic dynamics can preserve the spectral fluctuations that sustain it in the steady state.

\section{Connecting nonlocal magic to the capacity of entanglement}

Consider a pure state $|\Psi\rangle$ of $L$ qubits and a bipartition into a subsystem $A$ of $\LA\le L/2$ qubits and its complement $B$. Its Schmidt decomposition reads $|\Psi\rangle=\sum_{j=1}^{2^{\LA}}\sqrt{p_j}\,
|\psi_j\rangle_A\otimes|\varphi_j\rangle_B$, where the Schmidt probabilities are ordered as $p_1\ge p_2\ge\dots\ge0$ and satisfy $\sum_jp_j=1$.
Nonlocal magic is quantified by the minimum negative logarithm of the stabilizer fidelity~\cite{piroli2023stabilizer} over local changes of basis,
\begin{equation}
\begin{aligned}
\Dnl(|\Psi\rangle)=\min_{U_A,U_B}\bigl[&-\log_2 F_{\mathrm{STAB}}\bigl(
 (U_A\otimes U_B)|\Psi\rangle\bigr)\bigr],
\end{aligned}
\end{equation}
where $F_{\mathrm{STAB}}(|\psi\rangle)=\max_{|s\rangle\in\mathrm{STAB}_L}|\langle s|\psi\rangle|^2$ and $\mathrm{STAB}_L$ denotes the set of pure $L$-qubit stabilizer states. Here $U_A$ and $U_B$ are arbitrary unitaries acting within the two subsystems.
This optimization was recently solved exactly~\cite{sierant2026exact}, yielding
\begin{equation}
\begin{aligned}
\Dnl(|\Psi\rangle)&=-\log_2\max_{0\le k\le\LA}F_{2^k},\\
F_r&=\frac{1}{r}\left(\sum_{j=1}^{r}\sqrt{p_j}\right)^2.
\end{aligned}
\label{eq:exact}
\end{equation}

The maximization selects the number $k$ of Bell pairs whose Schmidt spectrum best matches that of $|\Psi\rangle$. The measure $\Dnl$ vanishes when this match is exact and diverges for families of states that universally embezzle entanglement~\cite{sierant2026exact}. Furthermore it bounds
tightly the nonlocal magic built from other monotones~\footnote{For example, minimizing the stabilizer R\'enyi entropy~\cite{hamma2022stabilizer,bittel2024stabilizer,piroli2023quantifying} over local unitaries defines an alternative measure~\cite{swingle2021conformal,oliviero2025gravitational,sierant2026spectral}, bounded above and below in terms of $\Dnl$~\cite{sierant2026exact}.}.

To characterize nonlocal magic in many-body systems, we consider the entanglement Hamiltonian $K=-\log_2\rho_A$, with eigenvalues $E_i=-\log_2p_i$. Its spectral distribution defines the \textit{entanglement energy spectrum}, $P(E)=\sum_i p_i\,\delta(E-E_i)$~\footnote{We use base-two logarithms. Factors of $\ln 2$ account for the natural-logarithm conventions of Refs.~\cite{qi2010entanglement,Boer19capacity,sierant2026exact}.}. We denote the cumulants of $P(E)$ by $\gamma_n$. The first two are $\gamma_1=\langle E\rangle_P\equiv S$, the entanglement entropy, and $\gamma_2=\langle E^2\rangle_P-\langle E\rangle_P^2\equiv C$, the capacity of entanglement~\cite{Boer19capacity}, named for its analogy with the heat capacity associated with $K$.
The full spectral dependence in Eq.~\eqref{eq:exact} simplifies when $P(E)$ is broad and varies slowly on the scale of one bit of entanglement energy; see Appendix~\ref{app:peak}. The cumulative Schmidt sums can then be approximated by integrals dominated by their upper energy cutoff $E_r=-\log_2p_r$, giving $F_r\simeq4P(E_r)/\ln2$ to leading order. Maximizing over $r=2^k$ selects the peak of the distribution at this order, yielding $\Dnl\simeq-\log_2(4P_{\max}/\ln2)$, with $P_{\max}=\max_E P(E)$. An approximately Gaussian distribution provides a further reduction to the capacity of entanglement; the finite-width construction is derived in Appendix~\ref{app:peak}.
Within the Gaussian approximation, we evaluate Eq.~\eqref{eq:exact} using a distribution with mean $S$ and variance $C$, obtaining
\begin{equation}
\begin{split}
F^{G}_{r}&=2^{-\frac{\ln 2}{4}C}\,
\frac{\Phi(z_r-\sigma/2)^{2}}{\Phi(z_r-\sigma)},
\\
z_r&=\sigma+\Phi^{-1}\!\Bigl(\frac{r}{d_G}\Bigr),
\label{eq:DG}
\end{split}
\end{equation}
where $\Phi$ is the standard normal cumulative distribution function and $d_G=2^{S+\frac{\ln2}{2}C}$ is the effective rank. The distribution has width $\sqrt{C}$ in our base-two convention, while $\sigma=\sqrt{C}\ln2$ expresses the same width in natural-logarithm units. For $r>d_G$, the expression continues as $F^{G}_{r}=2^{S+\frac{\ln2}{4}C}/r$. This gives the estimate $D_G(S,C)=-\log_2\max_kF^{G}_{2^k}$, which depends only on the first two cumulants.

The Gaussian peak height is $P_{\max}=(2\pi C)^{-1/2}$, so for $C\gg1$ the peak-height relation yields $\Dnl\simeq\frac12\log_2(\pi\ln^2\!2\,C/8)$. Near a flat spectrum,
the Gaussian expression instead gives $\Dnl\simeq(\ln2/4)\,C$. These limits describe broad spectra with large entanglement-energy fluctuations and nearly flat spectra with weak fluctuations, respectively, while $D_G(S,C)$ connects the two regimes. The approximation is controlled when the higher normalized cumulants satisfy $|\gamma_n|/C^{n/2}\ll1$ for $n\ge3$. In particular, when $C$ and the higher cumulants are extensive, these ratios scale as $\LA^{1-n/2}$ and vanish with increasing subsystem size, making the approximation asymptotically exact in the corresponding many-body scaling limit.

These insights provide a framework for predicting the dynamics of nonlocal magic from the entanglement entropy and capacity. We consider a chain of $L$ qubits initialized in a state $|\Psi_0\rangle=|\psi_0^A\rangle\otimes|\psi_0^B\rangle$ that is unentangled across the chosen bipartition. Since $\Dnl(0)=0$, all subsequent nonlocal magic is generated by the evolution $|\Psi_t\rangle=U_t|\Psi_0\rangle$. Below, we compare $\Dnl(t)$ evaluated exactly from the Schmidt spectrum through Eq.~\eqref{eq:exact} with the estimate $D_G(S(t),C(t))$. This comparison shows that the first two cumulants capture the qualitative behavior of the nonlocal magic across the models studied and yield quantitative agreement in the controlled scaling regimes identified below.

\section{Ergodic dynamics without conservation laws}

We first study nonlocal magic in a brickwall random quantum circuit with open boundary conditions~\cite{vijay2023random}. Its evolution is $U_t=U^{(t)}\cdots U^{(1)}$, with the circuit depth $t$ serving as time. Successive layers alternate between even and odd bonds, $U^{(2m)}=\bigotimes_i u_{2i,2i+1}$ and $U^{(2m+1)}=\bigotimes_i u_{2i-1,2i}$, with each two-qubit gate drawn independently from the Haar measure on $\mathrm{U}(4)$. We initialize $|\Psi_0\rangle=|0\rangle^{\otimes L}$ and average over at least $2000$ circuit realizations.

\begin{figure}[!t]
\centering
\includegraphics[width=0.96\columnwidth
]{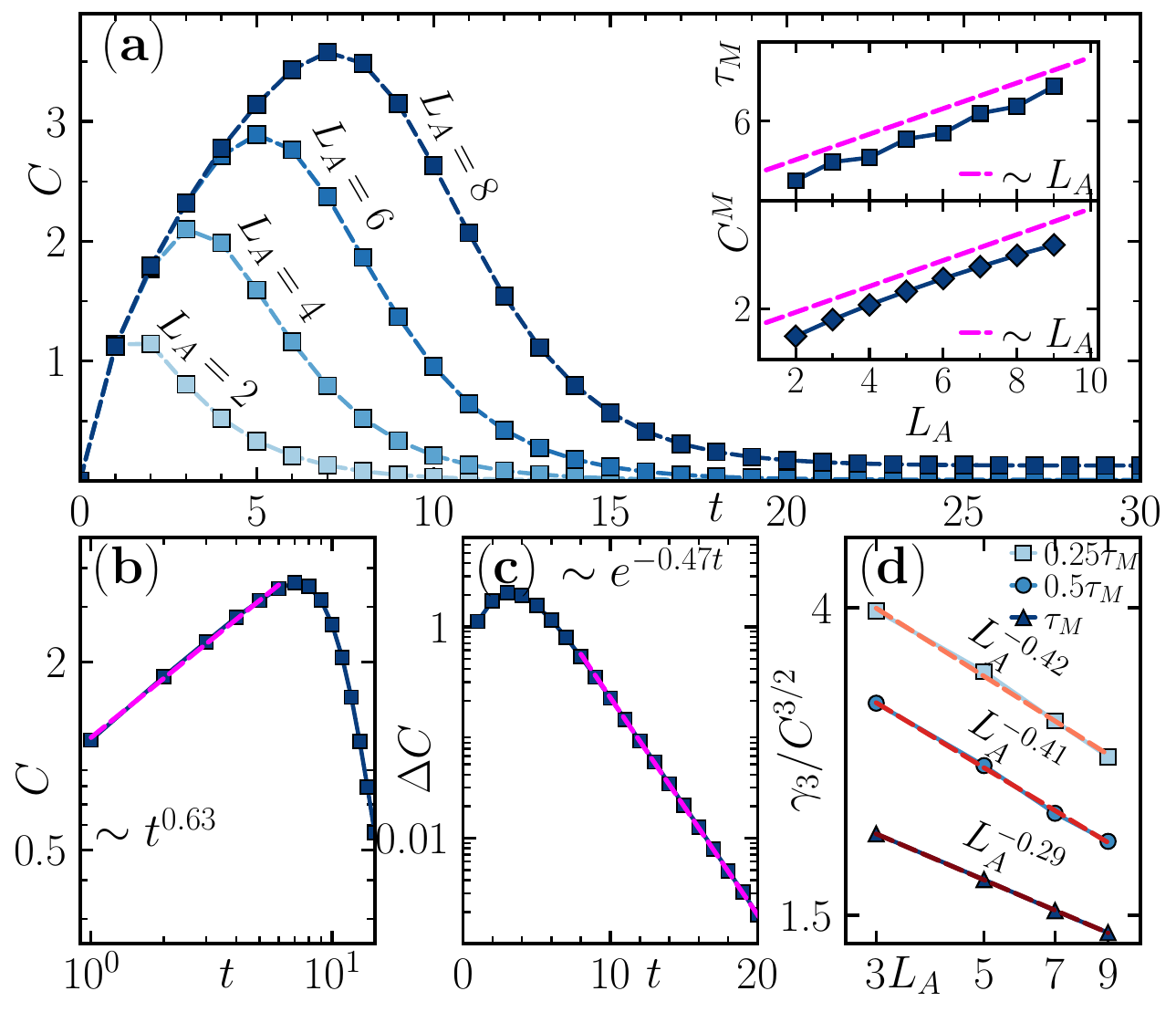}
\caption{\textbf{Capacity of entanglement in the Haar brickwall circuit.}
(a)~Rise--peak--fall profile of $C(t)$ for $\LA=2,4,6,8$. Insets show the linear scaling of the peak time $\tauM$ and peak value $C^M$ with $\LA$.
(b)~Algebraic early-time growth, fitted by $C\propto t^{0.63}$ over the accessible window.
(c)~Exponential relaxation of $\Delta C=C-C(\infty)$, with the same rate as $\Delta\Dnl$ in Fig.~\ref{fig:ergodic}.
(d)~Normalized third cumulant $\gamma_3/C^{3/2}$ versus $\LA$ at $t=0.25\tauM$, $0.5\tauM$, and $\tauM$, showing an algebraic decrease.
All data are for $L=20$. Markers denote exact state-vector results and dashed lines denote fits.}
\label{fig:capacity}
\end{figure}

We first examine the capacity, which probes the nonuniformity of the entanglement spectrum~\cite{karjula2026ebbsflowsquantumlearning}. Figure~\ref{fig:capacity}(a) shows that it follows a rise--peak--fall profile, even as the entanglement entropy grows towards saturation. Through Eq.~\eqref{eq:DG}, this evolution of spectral fluctuations translates into predictions for both the profile and characteristic scales of nonlocal magic.

During the initial growth regime, $C$ increases algebraically in time [Fig.~\ref{fig:capacity}(b)]. For sufficiently large subsystems, this growth opens a scaling window $1\ll t\ll\tauM$ in which $C(t)\gg1$. The large-capacity relation $\Dnl\simeq\tfrac12\log_2(\pi\ln^2\!2\,C/8)$ then converts the algebraic growth of $C$ into logarithmic growth of nonlocal magic, $\Dnl\propto\log_2t$. The capacity reaches an extensive maximum $C^M\propto\LA$ at $\tauM\propto\LA$ [insets of Fig.~\ref{fig:capacity}(a)], implying a nonlocal-magic peak whose time grows linearly and whose height grows logarithmically with the subsystem size $L_A$.

At late times, the excess capacity $\Delta C=C-C(\infty)$ decays exponentially [Fig.~\ref{fig:capacity}(c)]. For sufficiently unbalanced cuts, the spectrum approaches the near-flat regime, where $\Dnl\simeq(\ln2/4)\,C$, so that $\Delta\Dnl=\Dnl-\Dnl(\infty)$ inherits the exponential relaxation of $\Delta C$. The stationary value follows from the entanglement spectrum of a Haar-random state. For strongly unbalanced cuts, $\Delta=L-2\LA\gg1$, the large-dimension limit gives $C(\infty)\simeq2^{-\Delta}/\ln^2\!2$ and hence $\Dnl(\infty)\equiv D_\Delta\simeq2^{-\Delta}/(4\ln2)$~\cite{sierant2026exact}. At the balanced cut, the full spectral result instead gives the finite values $C(\infty)\simeq1.12$ and $D_0\simeq0.244$.

\begin{figure}[t!]
\centering
\includegraphics[width=\columnwidth]{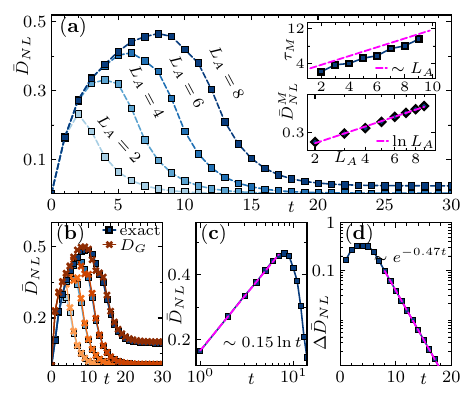}
\caption{\textbf{Nonlocal magic in the Haar-random brickwall circuit.}
(a)~Rise--peak--fall profile of $\Dnl(t)$ for $\LA=2,4,6,8$. Insets show the peak-time scaling $\tauM\propto\LA$ and peak-height scaling $\DM\propto\log_2\LA$.
(b)~Comparison of the Gaussian estimate $D_G(S,C)$ with $\Dnl$ evaluated from the exact Schmidt spectrum.
(c)~Initial logarithmic growth.
(d)~Exponential relaxation of $\Delta\Dnl=\Dnl-\Dnl(\infty)$, with a rate independent of $\LA$.
All data are for $L=20$. Markers denote exact state-vector results and magenta dashed lines denote fits.}
\label{fig:ergodic}
\end{figure}

Exact evaluation of $\Dnl$ from the Schmidt spectrum confirms these predictions [Fig.~\ref{fig:ergodic}]. The rise--peak--fall profile in panel (a) resembles that found for local resource measures in random circuits~\cite{Aditya2026growth,turkeshi2026coherence}. Its insets establish the linear peak-time and logarithmic peak-height scalings, while panels (c) and (d) show logarithmic growth and exponential relaxation, respectively. The relaxation rate is independent of $\LA$ and agrees with that of the capacity. Beyond these scaling relations, panel (b) directly compares $D_G(S,C)$ with the exact $\Dnl$: their agreement improves with increasing $\LA$. This improvement is consistent with the algebraic decrease of the normalized third cumulant $\gamma_3/C^{3/2}$ at fixed fractions of the peak time [Fig.~\ref{fig:capacity}(d)], supporting the Gaussian description over the accessible sizes.

Beyond random circuits, we also consider a chaotic Floquet model, the kicked Ising model, which exhibits the same rise--peak--fall profile and characteristic scalings. Its definition, simulation protocol, and numerical results are presented in Appendix~\ref{app:kim}.

\section{Ergodic dynamics with energy conservation}

\label{sec:mfim}

We now turn to non-integrable Hamiltonian dynamics, focusing on the mixed-field Ising model (MFIM)~\cite{huse2013ballistic}, $$H_{\rm MFIM}=b\sum_jX_j+h\sum_jZ_j+\sum_jZ_jZ_{j+1},$$ with open boundary conditions, transverse field $b=(\sqrt5+5)/8$, and longitudinal field $h=(\sqrt5+1)/4$. Continuous evolution under $H_{\rm MFIM}$ conserves energy. We average over $200$ random $z$-basis product states satisfying $|E_0-\bar E|/(E_{\max}-E_{\min})\le0.05$, where $E_0=\langle\Psi_0|H_{\rm MFIM}|\Psi_0\rangle$ and $\bar E$ is the center of the many-body spectrum~\cite{sierant2025anticoncentration}.

\begin{figure}[!htbp]
\centering
\includegraphics[width=\columnwidth]{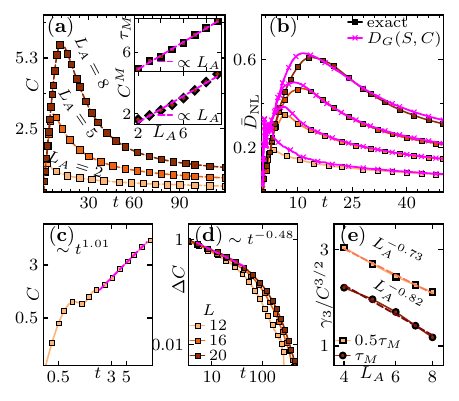}
\caption{\textbf{Capacity of entanglement in the mixed-field Ising model.}
(a)~$C(t)$ for $\LA=2,5,8$ at $L=20$. Insets show the linear scaling of the peak time $\tauM$ and peak value $C^M$ with $\LA$.
(b)~Comparison of the exact $\Dnl$ with the Gaussian estimate $D_G(S,C)$ for $\LA=2,4,6,8$.
(c)~Approximately linear early-time growth, fitted by $C\propto t^{1.01}$.
(d)~Algebraic relaxation of $\Delta C=C-C(\infty)$ at $\LA=2$ for $L=12,16,20$, fitted by $\Delta C\propto t^{-0.48}$. The stationary capacity is averaged over $800\le t\le1000$.
(e)~Normalized third cumulant $\gamma_3/C^{3/2}$ versus $\LA$ at $t=0.5\tauM$ and $\tauM$.
Markers denote exact state-vector results and dashed lines denote fits.}
\label{fig:capacity_mfim}
\end{figure}

The capacity again follows a rise--peak--fall profile while the entropy grows towards saturation [Fig.~\ref{fig:capacity_mfim}(a)]. Its initial growth is approximately linear in time [Fig.~\ref{fig:capacity_mfim}(c)], opening, for sufficiently large subsystems, a window $1\ll t\ll\tauM$ in which $C(t)\gg1$. The large-capacity branch of Eq.~\eqref{eq:DG} therefore predicts logarithmic growth of nonlocal magic, $\Dnl\propto\log_2t$. The capacity reaches an extensive maximum $C^M\propto\LA$ at $\tauM\propto\LA$ [insets of Fig.~\ref{fig:capacity_mfim}(a)], suggesting the same linear peak-time and logarithmic peak-height scalings as in the circuit. Energy conservation changes the subsequent relaxation: at fixed $\LA$ and increasing $L$, $\Delta C=C-C(\infty)$ decays algebraically [Fig.~\ref{fig:capacity_mfim}(d)]. In the near-flat regime, the relation $\Dnl\simeq(\ln2/4)\,C$ then predicts algebraic relaxation of $\Delta\Dnl$ as well.

\begin{figure}[!htbp]
\centering
\includegraphics[width=\columnwidth]{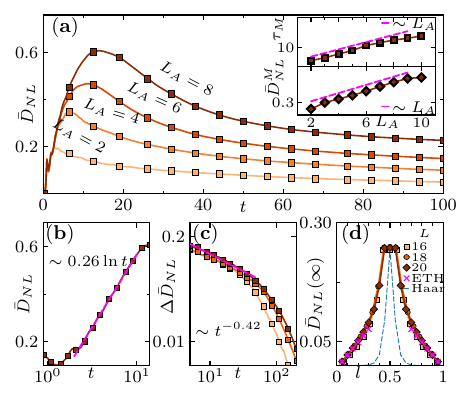}
\caption{\textbf{Nonlocal magic in the mixed-field Ising model.}
(a)~Rise--peak--fall profile of $\Dnl(t)$. Insets show a peak time linear in $\LA$ and a peak height growing approximately linearly over the accessible sizes.
(b)~Initial logarithmic growth.
(c)~Algebraic relaxation of $\Delta\Dnl=\Dnl-\Dnl(\infty)$, fitted by $\Delta\Dnl\propto t^{-0.42}$.
(d)~Stationary nonlocal magic versus the subsystem fraction $\ell=\LA/L$, compared with the thermal prediction of Eq.~\eqref{eq:eth} (crosses) and the balanced-cut Haar value $D_0\simeq0.244$ (dashed line).
We use $L=20$ in panels (a)--(b) and $L=16,18,20$ in panels (c)--(d). Markers denote exact state-vector results and magenta dashed lines denote fits.}
\label{fig:ergodic_mfim}
\end{figure}

The exact nonlocal magic confirms this qualitative picture:
it exhibits a rise--peak--fall profile with $\tauM\propto\LA$ [Fig.~\ref{fig:ergodic_mfim}(a)], logarithmic initial growth [Fig.~\ref{fig:ergodic_mfim}(b)], and algebraic relaxation [Fig.~\ref{fig:ergodic_mfim}(c)]. The main departure from the capacity-based prediction concerns the peak height, which grows faster than $\log_2\LA$, approximately linearly over the accessible sizes, with slower growth near the balanced cut [insets of Fig.~\ref{fig:ergodic_mfim}(a)]. We interpret this behavior as a preasymptotic effect of the limited subsystem sizes.

The comparison with $D_G(S,C)$ further clarifies the scope of the spectral approximation. Although the normalized third cumulant decreases with $\LA$ [Fig.~\ref{fig:capacity_mfim}(e)], the discrepancy between $D_G$ and $\Dnl$ does not decrease systematically, either near the peak or during the subsequent relaxation. Thus, decreasing skewness alone does not establish convergence to a Gaussian spectrum. Over the sizes studied, $D_G$ follows the exact dynamics as an upper envelope [Fig.~\ref{fig:capacity_mfim}(b)], capturing its qualitative evolution without the improving quantitative agreement observed in the circuit.

Energy conservation also changes the stationary spectrum. For a small subsystem, $\LA\ll L$, the eigenstate thermalization hypothesis (ETH)~\cite{deutsch2018eigenstate,srednicki1994chaos,olshanii2008thermalization,rigol2016quantum,eisert2016equilibration,kurchan2022eigenstate,hastings2011strong,fagotti2014conservation,hamma2025entropy,hauke2025deep} motivates the thermal approximation $\rho_A(\infty)\simeq e^{-\beta H_A}/Z_A$, where $H_A$ contains the Hamiltonian terms within $A$ and $Z_A=\mathrm{Tr}\,e^{-\beta H_A}$. The inverse temperature $\beta$ is fixed by matching the thermal energy density to that of the initial state. At high temperature, this gives $\beta\simeq-E_0/V_L$, with $V_L=(L-1)+L(h^2+b^2)$ the infinite-temperature variance of the full Hamiltonian.

The thermal form makes the stationary capacity directly accessible: since $K=(\beta H_A+\ln Z_A)/\ln2$, we have $C(\infty)\simeq\beta^2\mathrm{Var}_{\beta}(H_A)/\ln^2\!2$. Expanding the thermal variance around infinite temperature and applying the near-flat relation for $\Dnl$ yields
\begin{equation}
C(\infty)\simeq\frac{\beta^2V_A}{\ln^2\!2},\qquad
\Dnl(\infty)\simeq\frac{\beta^2V_A}{4\ln2},
\label{eq:eth}
\end{equation}
where $V_A=(\LA-1)+\LA(h^2+b^2)$. The first correction involves the third infinite-temperature cumulant of $H_A$; the derivation and higher-order terms are given in Appendix~\ref{app:eth}. Since $V_A$ is extensive, Eq.~\eqref{eq:eth} predicts an approximately linear increase of the stationary nonlocal magic with $\LA$, within the small-subsystem and near-flat regimes.

Equation~\eqref{eq:eth} agrees well with the stationary values for smaller subsystems [Fig.~\ref{fig:ergodic_mfim}(d)]. The agreement deteriorates towards the balanced cut, where the small-subsystem thermal approximation no longer applies and the numerical values approach the Haar result $D_0\simeq0.244$. For the quenches studied, this behavior is consistent with an $O(1)$ stationary nonlocal magic. Energy conservation therefore slows its relaxation and modifies its stationary value, while preserving the transient production and subsequent loss of nonlocal magic observed in the circuit and Floquet models.

\begin{figure}[!htbp]
\centering
\includegraphics[width=0.96\columnwidth]{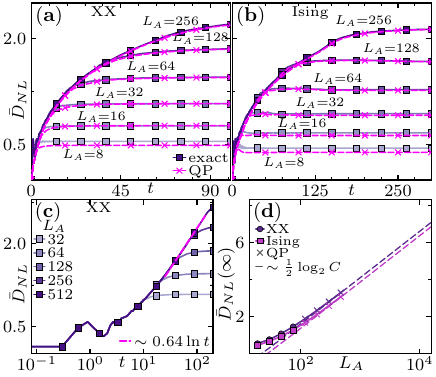}
\caption{\textbf{Nonlocal magic in free-fermionic Hamiltonian quenches.}
(a)--(b)~Growth--saturation profiles of $\Dnl(t)$ in the XX and transverse-field Ising chains, comparing exact results (squares) with the quasiparticle prediction (crosses).
(c)~Initial logarithmic growth, $\Dnl\propto\log_2t$.
(d)~Stationary values, obtained from long-time averages for $\LA\le64$ and from the quasiparticle prediction for $\LA>64$, approach the asymptotic law $\tfrac12\log_2(\pi\ln^2\!2\,C/8)$.
We use $L=4096$ sites and subsystem sizes $\LA=8$--$512$.}
\label{fig:gauss}
\end{figure}

\section{Integrable dynamics}

We now turn to integrable Hamiltonians and exactly tractable free-fermionic dynamics. We develop a quasiparticle picture for the production and transport of nonlocal magic, derive and quantitatively test its predictions in free-fermionic chains, and extend the stationary analysis to interacting integrable models through the thermodynamic Bethe ansatz. Finally, random matchgate circuits allow us to explore the same connection between nonlocal magic and capacity beyond ballistic quasiparticle propagation.

\subsection{Free-fermionic Hamiltonians}

We consider the XX chain, $H_{\rm XX}=-\tfrac12\sum_j(\sigma^x_j\sigma^x_{j+1}+\sigma^y_j\sigma^y_{j+1})$, and the transverse-field Ising chain, $H_{\rm TFI}=-\sum_j\sigma^x_j\sigma^x_{j+1}-h\sum_j\sigma^z_j$, at $h=0.5$. Both models are defined on periodic spin chains and quenched from the dimer state $|\Psi_0\rangle=\bigotimes_j\tfrac{1}{\sqrt2}(|{\uparrow\downarrow}\rangle-|{\downarrow\uparrow}\rangle)_{2j,2j+1}$, with the block boundaries chosen between dimers. The initial state has $N=L/2$ Jordan--Wigner fermions, fixing a fermion parity conserved by both Hamiltonians. The fermions therefore obey periodic boundary conditions for $L/2$ odd and antiperiodic ones for $L/2$ even.

For these states, the reduced density matrix factorizes in its canonical fermionic basis as $\rho_A=\bigotimes_{j=1}^{\LA}\mathrm{diag}(n_j,1-n_j)$. The entanglement energy is consequently a sum of independent two-valued contributions, making every cumulant of $P(E)$ additive. In particular,
\begin{equation}
S=\sum_{j=1}^{\LA}h(n_j),\qquad
C=\sum_{j=1}^{\LA}c(n_j),
\label{eq:gaussSC}
\end{equation}
where $h(n)=-n\log_2n-(1-n)\log_2(1-n)$ and $c(n)=n(1-n)\log_2^2[n/(1-n)]$. The occupations $n_j$, and hence $S$ and $C$, follow efficiently from the covariance matrix~\cite{Peschel2012Entanglement,Peschel2004ReducedDensityMatrix}, giving access to the spectral prediction well beyond the sizes accessible to full state-vector evolution. When $m$ modes contribute an extensive capacity, $C\propto m$, with higher cumulants of order $m$, their normalized values vanish as $|\gamma_3|/C^{3/2}=O(m^{-1/2})$ and $|\gamma_4|/C^2=O(m^{-1})$. Together with a smooth spectral limit, this makes the Gaussian estimate $D_G(S,C)$ asymptotically controlled.

We develop the quasiparticle picture of nonlocal magic by expressing both $S$ and $C$ in terms of the excitations produced by the quench. In the pair description, quasiparticles contribute to the block spectrum when they are shared between $A$ and its complement. The same geometric weight therefore enters the entropy and capacity~\cite{cardy2005evolution,calabrese2008evolution},
\begin{equation}
\begin{split}
    S_{\rm QP}&=\int\!\frac{dk}{2\pi}\,w_k(t)\,h(n_k),\\
C_{\rm QP}&=\int\!\frac{dk}{2\pi}\,w_k(t)\,c(n_k),
\label{eq:qp}
\end{split}
\end{equation}
where $n_k$ are the postquench mode occupations, $v_k$ their group velocities, and $w_k(t)=\min(\LA,2|v_k|t)$ encodes the ballistic separation of each pair. Substitution into Eq.~\eqref{eq:DG} then yields the dynamical prediction $\Dnl(t)\simeq D_G(S_{\rm QP},C_{\rm QP})$.

Before saturation, $C_{\rm QP}$ grows linearly in time. For sufficiently large blocks, there is consequently a window beyond microscopic times in which $C\gg1$, and the large-capacity relation gives $\Dnl\propto\log_2t$. This explains why logarithmic growth occurs in both free-fermionic and ergodic dynamics: in either case, algebraically growing entanglement-energy fluctuations are converted into logarithmic nonlocal magic. Their late-time behavior differs because the quasiparticle occupations retain a nonzero capacity density. In the stationary thermodynamic limit, $w_k\to\LA$ and $C(\infty)=\bar c\,\LA$, with $\bar c=\int\frac{dk}{2\pi}c(n_k)$, giving

\begin{equation}
\Dnl(\infty)=\tfrac12\log_2\LA
+\tfrac12\log_2\frac{\pi\ln^2\!2\,\bar c}{8}
+O(\LA^{-1}).
\label{eq:gaussinf}
\end{equation}

The quench enters through the capacity density and hence the intercept, while the coefficient of $\log_2\LA$ is fixed.

This stationary law requires $\bar c>0$. For example, a N\'eel-state quench in the XX chain has $n_k=\tfrac12$ for every $k$, so that $\bar c=0$ and the block relaxes to the maximally mixed state. In that case, $\Dnl$ decays as $t^{-1}$ instead of approaching a logarithmically growing plateau; see Appendix~\ref{app:flat_occupations}. Persistent nonlocal magic thus depends on the spectral fluctuations retained by the stationary state.

Our numerical results in Fig.~\ref{fig:gauss} confirm the quasiparticle predictions. Panels (a)--(b) show logarithmic growth followed by saturation, with quantitative agreement beyond the microscopic transient. The fastest quasiparticles set the onset of the saturation crossover at $t\sim\LA/(2v_{\max})$, while slower modes govern the subsequent approach to the plateau. Panel (c) resolves the logarithmic growth, and panel (d) approaches the stationary scaling $\Dnl(\infty)\sim\tfrac12\log_2\LA$ of Eq.~\eqref{eq:gaussinf}. For these quenches, the extensive stationary capacity therefore sustains a diverging nonlocal magic, making the resulting families of states universal entanglement embezzlers.

\subsection{Interacting integrable models}

\label{sec:interacting}

The stationary mechanism also extends to interacting systems. We use the thermodynamic Bethe ansatz (TBA) to compute the entropy and capacity densities of the generalized Gibbs ensemble selected by the quench~\cite{olshanii2007relaxation,rigol2016generalized}. In the stationary thermodynamic limit, the block is described, up to boundary corrections, by $\rho_A\simeq e^{-\mathcal K_A}/Z_A$, where $\mathcal K_A=\sum_j\beta_jQ_{j,A}$ is the generalized Gibbs generator built from the conserved charges restricted to $A$. We use natural logarithms in the TBA construction, so the entanglement Hamiltonian defined previously is $K=(\mathcal K_A+\ln Z_A)/\ln2$.

In the string description, the stationary macrostate is specified by particle and hole densities $\rho_n(\lambda)$ and $\rho_n^h(\lambda)$, with total density $\rho_n^t=\rho_n+\rho_n^h$ and filling $\vartheta_n=\rho_n/\rho_n^t$. They obey the Bethe--Takahashi equations~\cite{takahashi2001simplification} $\rho_n^t=a_n-\sum_m a_{nm}*\rho_m$, where $a_n$ is the bare density of states and $a_{nm}$ is the scattering kernel. Here $*$ is the convolution in rapidity, $(f*g)(\lambda)=\int d\lambda'\,f(\lambda-\lambda')g(\lambda')$, so that each string species feels every other one integrated over the rapidities it occupies. Writing $\eta_n=\vartheta_n^{-1}-1$, the TBA equations read
\begin{equation}
\ln\eta_n=w_n+\sum_m a_{nm}*\ln(1+\eta_m^{-1}),
\label{eq:tba}
\end{equation}
with the driving terms $w_n$ fixed by the quench-selected ensemble. The entropy density is the Yang--Yang entropy,
\begin{equation}
\bar s=\sum_n\int d\lambda\,\rho_n^t\,h(\vartheta_n),
\label{eq:sgge}
\end{equation}
where $h$ is defined in Eq.~\eqref{eq:gaussSC}, so that $S\simeq\LA\bar s$.

To obtain the capacity, we introduce the tilted partition function $Z_A(s)=\mathrm{Tr}\,e^{-s\mathcal K_A}$. Its second logarithmic derivative gives the variance of $\mathcal K_A$, and hence $C=(\ln2)^{-2}\partial_s^2\ln Z_A(s)|_{s=1}$. The tilt changes the TBA driving terms as $w_n\to sw_n$. Differentiating Eq.~\eqref{eq:tba} gives $\partial_s\ln\eta_n|_{s=1}=w_n^{\rm dr}$, where the dressed driving terms satisfy $w_n^{\rm dr}=w_n-\sum_m a_{nm}*(\vartheta_m w_m^{\rm dr})$. Evaluating the corresponding response of the generalized free-energy density, $\lim_{\LA\to\infty}\LA^{-1}\ln Z_A(s)=\sum_n\int d\lambda\,a_n\ln[1+\eta_n(s)^{-1}]$, yields $C\simeq\LA\bar c$, with

\begin{equation}
\bar c=\frac{1}{\ln^2\!2}\sum_n\int d\lambda\,
\rho_n^t\,\vartheta_n(1-\vartheta_n)
\bigl[w_n^{\rm dr}\bigr]^2.
\label{eq:cgge}
\end{equation}
This is the TBA expression for a generalized charge susceptibility: $\vartheta_n(1-\vartheta_n)$ weights particle--hole fluctuations, while the dressed driving term accounts for the interactions.

Higher derivatives of the same partition function generate the higher cumulants of the entanglement energies. When $\ln Z_A(s)$ has a smooth extensive thermodynamic limit, every fixed-order cumulant is $O(\LA)$. For $\bar c>0$, their normalized values therefore decrease as $|\gamma_p|/C^{p/2}=O(\LA^{1-p/2})$ for $p\ge3$. If the entanglement-energy distribution also admits a regular local saddle-point expansion, the large-capacity relation gives
\begin{equation}
\Dnl(\infty)=\tfrac12\log_2\LA
+\tfrac12\log_2\frac{\pi\ln^2\!2\,\bar c}{8}
+O(\LA^{-1}).
\label{eq:tbapred}
\end{equation}
The same logarithmic law thus survives interactions under these conditions. The quench and interaction strength enter through the susceptibility $\bar c$, changing the intercept while leaving the leading coefficient unchanged. These interacting integrable steady-state families consequently support universal entanglement embezzlement as well.

A direct numerical test of Eq.~\eqref{eq:tbapred} remains challenging. Resolving its asymptotic regime requires blocks with $\LA\bar c\gg1$, while exact diagonalization is restricted to small systems and tensor-network evolution towards the stationary state is limited by the entanglement growth following a global quench. Numerical verification of the interacting prediction therefore remains for future work.

\subsection{Beyond the quasiparticle picture: random matchgate circuits}

\label{sec:matchgate}
The capacity framework also applies when the ballistic quasiparticle construction is unavailable. We consider random free-fermionic matchgate circuits in brickwall geometry~\cite{lumiamatchgate25,dias2021diffusiveoperatorspreadingrandom,sierant2026theorymatchgatecommutant,trigueros2026unitarydesignsdopedmatchgate}, initialized in $|\Psi_0\rangle=|0\rangle^{\otimes L}$. Each two-site gate is an independently sampled free-fermionic unitary: a parity-preserving matchgate $G(A,B)$ with $A,B\in\mathrm{U}(2)$ and $\det A=\det B$. Such a gate rotates the four local Majorana operators by $O\in\mathrm{SO}(4)$, evolving their covariance matrix as $\Gamma\to O\Gamma O^T$, where $\Gamma_{ab}=\tfrac{i}{2}\langle[\chi_a,\chi_b]\rangle$. The dynamics therefore remains exactly tractable at the covariance-matrix level, while temporal randomness produces diffusive single-particle transport.

Equation~\eqref{eq:gaussSC} continues to determine $S(t)$ and $C(t)$ directly from the evolving mode occupations. The large-capacity relation still converts algebraic capacity growth into logarithmic growth of $\Dnl$, but diffusion sets a saturation time $\tauM\propto\LA^2$. The resulting growth--saturation profile therefore persists without ballistic propagation. At fixed diffusive time $u=t/\LA^2$, a macroscopic number of contributing modes provides the same mechanism for suppressing higher normalized cumulants as in the free-fermionic Hamiltonians. In particular, the central-limit prediction for the skewness is $|\gamma_3|/C^{3/2}\propto\LA^{-1/2}$.

\begin{figure}[!t]
\centering
\includegraphics[width=\columnwidth]{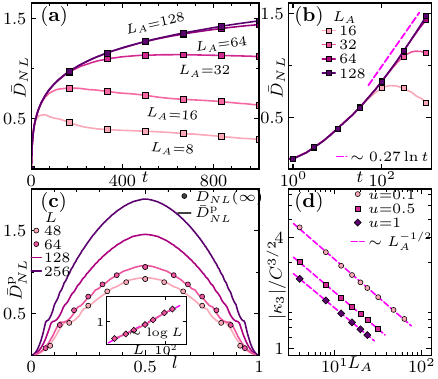}
\caption{\textbf{Nonlocal magic beyond ballistic quasiparticle propagation.}
(a)~Growth--saturation profile of $\Dnl(t)$ in random matchgate circuits at $L=256$ for $\LA=8,16,32,64,128$.
(b)~The same data on a logarithmic time axis, resolving the initial logarithmic growth.
(c)~Stationary Page curve versus $\ell=\LA/L$. Markers show time-averaged stationary values for $L=48,64$; lines show the analytical prediction for $L=48,64,128,256$. The inset displays the balanced-cut scaling $\tfrac12\log_2L$.
(d)~Normalized third cumulant $|\gamma_3|/C^{3/2}$ versus $\LA$ at fixed diffusive times $u=t/\LA^2=0.1,0.5,1$, following the central-limit scaling $\LA^{-1/2}$.
Numerical data are obtained from covariance-matrix evolution and averaged over circuit realizations; magenta dashed lines denote fits.}
\label{fig:matchgate}
\end{figure}

The stationary ensemble consists of random pure free-fermion states.
  For the smaller subsystem, with $\ell=\LA/L\le\tfrac12$, the mode occupations
  have the limiting density~\cite{bianchi2021page},
  $\varrho_\ell(n)=\sqrt{(n_+-n)(n-n_-)}/[2\pi\ell\,n(1-n)]$,
  supported on $[n_-,n_+]$, where
  $n_\pm=\tfrac12\pm\sqrt{\ell(1-\ell)}$.
Averaging the single-mode functions in Eq.~\eqref{eq:gaussSC} gives $S=\LA\bar h(\ell)$ and $C=\LA\bar c(\ell)$, where $\bar h(\ell)=\int_{n_-}^{n_+}dn\,\varrho_\ell(n)h(n)$ and $\bar c(\ell)=\int_{n_-}^{n_+}dn\,\varrho_\ell(n)c(n)$. The Gaussian estimate $D_G(\LA\bar h(\ell),\LA\bar c(\ell))$ then yields the stationary nonlocal-magic Page curve; the derivation is given in Appendix~\ref{app:page_curve}.

At fixed nonzero $\ell$, the stationary capacity is extensive, giving a logarithmically growing Page curve whose maximum at the balanced cut scales as $\tfrac12\log_2L+O(1)$. For much smaller blocks, $\bar c(\ell)\simeq\ell/\ln^2\!2$, so that $C\simeq\LA^2/(L\ln^2\!2)$. In the near-flat regime $\LA^2/L\ll1$, the small-capacity branch instead gives $\Dnl(\infty)\simeq\LA^2/(4L\ln2)$. The same spectral description thus connects the small-cut behavior to the logarithmically growing values on macroscopic bipartitions.

Our numerical results confirm this picture [Fig.~\ref{fig:matchgate}]. Panels (a)--(b) show logarithmic growth followed by saturation, with no late-time decay. Panel (c) compares the stationary values at $L=48$ and $64$ with the predicted Page curves, which also provide predictions for larger systems. Its inset confirms the balanced-cut scaling $\Dnl(\infty)\sim\tfrac12\log_2L$. Finally, panel (d) shows the expected $\LA^{-1/2}$ decrease of the normalized third cumulant at fixed $u$, supporting the reduction to the first two cumulants in the diffusive scaling regime.

Random matchgate circuits demonstrate that sustained nonlocal magic does not require ballistic quasiparticles. Their stationary states retain extensive entanglement-energy fluctuations on macroscopic bipartitions, supporting universal embezzlement even when transport is diffusive. The capacity therefore identifies the spectral mechanism responsible for retention of nonlocal magic across coherent free-fermionic quenches, interacting integrable steady states, and temporally random free-fermionic evolution.

\section{Discussion and outlook}

We have shown that the spreading of nonlocal magic is governed by the capacity
of entanglement. The early-time growth is universal,
$\Dnl\sim\log_2 t$. At late times, ergodic dynamics displays a
rise--peak--fall profile with an $O(1)$ stationary value, whereas free-fermionic
dynamics saturates at $\Dnl(\infty)\propto\log_2\LA$. The framework extends to
interacting integrable models, where the thermodynamic Bethe ansatz yields the
two generalized-Gibbs-ensemble cumulants fixed by the quench
action~\cite{calabrese2017entanglement,calabrese2018entanglement,
rigol2016generalized,olshanii2007relaxation,mussardo2016introduction,
yoshimura2016emergent,fagotti2016transport}.

Operationally, our results link scrambling to entanglement embezzlement. A
family universally embezzles entanglement if and only if $\Dnl$ diverges along
it~\cite{hayden2003universal,gour2024complete,wilming2025critical,
sierant2026exact}. Counterintuitively, the strongest scramblers embezzle only
transiently: the flat spectra they produce destroy their catalytic power in the long time limit of the evolution.
Free-fermionic dynamics, despite its efficient classical simulability and restricted scrambling power, embezzles universally. Catalytic power is therefore set not by the amount of entanglement generated, but by the shape of the resulting spectrum, accessible in simulations and experiments~\cite{hamma2025experimental}.

Several questions remain open, and the entanglement capacity makes them analytically tractable. Determining whether non-unitary
ingredients of dynamics, such as measurement~\cite{Oshima2025Topology,Li2025MeasurementInduced,vijay2023random} and dissipation~\cite{Ashida2020NonHermitian,Kawabata2026NonHermitian,LeGal2023VolumeToArea,turkeshi2023entanglement} remove
nonlocal magic or reshape its
profile would be worth investigating. The same applies to
symmetries and ergodicity
breaking~\cite{zakrzewski2022challenges,abanin19MBLreview,Sierant2025MBL,
tirrito2025nonstabilizerness,Moudgalya_2022}. For systems with permutation
symmetry~\cite{Passarelli24perm} a single collective degree of freedom controls the spectrum, and a
closed formula for $\Dnl$ should follow. Resolving the asymptotic scaling of
the integrable prediction requires more involved numerical investigations, thus remains for future work. Finally, it remains to connect it to quantum
metrology~\cite{pappalardi2026assembling,podzie2026stabilizerness} and to the
information lattice~\cite{Kvorning2022TimeEvolution,Artiaco2024LITE,Artiaco2025Universal}. 
Exploring Mpemba-like effects~\cite{Ares2023Asymmetry, Ares2025MpembaReview,Turkeshi25mpemba,Aditya26mpemba,aditya2026higherordersymmetricquantummpemba} in nonlocal magic would also be worthwhile.

\begin{acknowledgments}

We thank B. Magni, D. Iannotti, R. Cioli and A. Hamma for collaborations on related topics and discussions. 
S. A. is funded by an Alexander von Humboldt fellowship.
X.T. acknowledges support from DFG Emmy Noether Programme proposal ``Digital Quantum Matter Out-of-Equilibrium'' No.~560726973, DFG under Germany's Excellence Strategy -- Cluster of Excellence Matter and Light for Quantum Computing (ML4Q) EXC~2004/2 -- 390534769, and DFG Collaborative Research Center (CRC)~183 Project No.~277101999 -- project B01.
P.S. acknowledges a fellowship within the ``Generaci\'on D'' initiative, Red.es, Ministerio para la Transformaci\'on Digital y de la Funci\'on P\'ublica, for talent attraction (C005/24-ED CV1), funded by the European Union NextGenerationEU funds, through PRTR.

{\it Note added--} While preparing this manuscript, we became aware of three independent works~\cite{zhang2026entanglementgrowthtransportschmidt,li2026nonlocalmagicmanybodylocalization,karjula2026universalentanglementembezzlementdivergent} exploring related directions. Where our results overlap, they are in agreement.
The relationship between the fermionic nonlocal magic in free fermionic systems and the capacity of entanglement is furthermore discussed in a recent independent work~\cite{magniiannotti}.

{\it Code and Data Availability.--} The code and the data
for our simulations will be publicly shared at publication.

\end{acknowledgments}


\appendix
\begin{widetext}

\section{Proof of the capacity law}%
\label{app:peak}
The main text asserts that the nonlocal magic of a regular entanglement
spectrum is fixed by the capacity alone,
\begin{equation}
\Dnl\simeq
\begin{cases}
\tfrac12\log_2\bigl(\pi\ln^{2}\!2\,C/8\bigr), & C\gg1,\\[3pt]
(\ln2/4)\,C, & C\ll1,
\end{cases}
\label{eq:caplaw}
\end{equation}
with the Gaussian closure interpolating between the two. We prove both branches
here.
 
We first prove a law that makes no reference to the
capacity at all: for any profile smooth on the scale of one bit, $\Dnl$ is the
logarithm of the height of $P$ at its mode. The capacity enters only afterwards,
once a shape is committed to, and it enters differently in the two limits. At
large capacity the profile is Gaussian and its mode height is $(2\pi C)^{-1/2}$.
At small capacity the profile is narrower than one bit, the smoothness
hypothesis fails outright, and the second branch has to be obtained from the
discrete spectrum instead. Appendix~\ref{app:bounds} says what survives when
neither hypothesis holds.
 
\subsection{The peak-height law}
 
A \emph{prefix} of the sorted spectrum is the set of Schmidt values of entanglement
energy below some $E_r$. Using $1=2^{E_i}p_i$ and $\sqrt{p_i}=2^{E_i/2}p_i$ to
rewrite its size and its amplitude sum as integrals against the Born profile, we
obtain the counting and the amplitude functions
\begin{eqnarray}
R(E)&=&\#\{i:E_i\le E\}=\int^{E}\!2^{E'}P(E')\,dE',\nonumber\\
A(E)&=&\sum_{E_i\le E}\sqrt{p_i}=\int^{E}\!2^{E'/2}P(E')\,dE',
\label{eq:RA}
\end{eqnarray}
so that the prefix fidelity is $F_r:=A(E_r)^{2}/R(E_r)$ with $R(E_r)=r$.
Explicitly, $R$, $A$ and $P$ are one measure at the three tilts $2^{E}$,
$2^{E/2}$ and $1$.
 
Smoothness here means that $P$ varies slowly across one bit of entanglement
energy, the spacing between consecutive dyadic ranks. Since $P$ has width
$\sqrt{C}$, the requirement is $C\gg1$: the capacity alone decides whether the
regime applies.
 
Both tilts weight the prefix towards its upper endpoint over a range of order
one bit, so that only the behavior of $P$ near $E_r$ matters. Substituting
$E'=E_r-u/\ln2$ and using $\int_0^\infty\!e^{-u}u^{n}du=n!$ together with
$\int_0^\infty\!e^{-u/2}u^{n}du=2^{n+1}n!$, the two integrals become series in
the derivatives of $P$ at $E_r$,
\begin{eqnarray}
R&=&\frac{2^{E_r}}{\ln2}
\Bigl[P-\frac{P'}{\ln2}+\frac{P''}{\ln^{2}\!2}-\dots\Bigr],\nonumber\\
A&=&\frac{2^{E_r/2+1}}{\ln2}
\Bigl[P-\frac{2P'}{\ln2}+\frac{4P''}{\ln^{2}\!2}-\dots\Bigr],
\label{eq:RAexp}
\end{eqnarray}
whose ratio is
\begin{equation}
F_r=\frac{A^{2}}{R}=\frac{4P(E_r)}{\ln2}
\Bigl[1-\frac{3P'}{P\ln2}+\frac{7P''}{P\ln^{2}\!2}
+O\bigl(P'^{2}/P^{2}\bigr)\Bigr].
\label{eq:Fexp}
\end{equation}
Every derivative costs a factor $1/(\sqrt{C}\ln2)$, so the series is controlled
precisely in the regime just defined.
 
The bound in Eq.~\eqref{eq:Fexp} is stationary where $P'=0$, that is at the
mode of $P$, and there
\begin{equation}
\max_{0\le k\le\LA}F_{2^k}\simeq\frac{4P_{\max}}{\ln2},
\qquad
\Dnl\simeq-\log_2\frac{4P_{\max}}{\ln2},
\label{eq:peak}
\end{equation}
which is the peak-height law. The corrections are of relative order $1/C$: the
first derivative vanishes at the mode, leaving the curvature term
$7P''/(P\ln^{2}\!2)=-7/(C\ln^{2}\!2)$ as the leading one. Restricting $r$ to
powers of two costs no more. The dyadic grid locates the mode to within one
bit, and a displacement $\delta E$ from a stationary point lowers $\ln P$ by
$\delta E^{2}/2C$, again $O(1/C)$ for $\delta E=O(1)$, because $P$ varies on the
scale $\sqrt{C}$ and not on the scale of the octave (here and below, an \textit{octave} is a window of ranks differing by at most a factor of two) spacing.
 
The mean drops out. Translating $P$ along the energy axis changes the
entanglement entropy and leaves $P_{\max}$, hence $\Dnl$, untouched: the shape
of the profile fixes the nonlocal magic, not the entanglement it carries. 
We emphasize that Eq.~\eqref{eq:peak} holds for
every smooth profile, independently of its width.
 
\subsection{Large capacity regime}
 
The capacity enters when a shape is fixed. A profile with suppressed normalized
cumulants is Gaussian, of variance $C$ and mode height
$P_{\max}=(2\pi C)^{-1/2}$, and Eq.~\eqref{eq:peak} returns the first line of
Eq.~\eqref{eq:caplaw},
\begin{equation}
\Dnl\simeq\tfrac12\log_2\frac{\pi\ln^{2}\!2\,C}{8},
\qquad(C\gg1,\ \gamma_3\to0).
\label{eq:peakgauss}
\end{equation}
 
One order further fixes the first correction. Put $E_r=S+y$ in
Eq.~\eqref{eq:Fexp} and use $P'/P=-y/C$, $P''/P=y^{2}/C^{2}-1/C$, so that
$\ln F_r=\mathrm{const}-y^{2}/2C+3y/(C\ln2)-7/(C\ln^{2}\!2)+O(C^{-2})$. The
maximum moves to $y^\star=3/\ln2$ bits, above the mean. Displacement
contributes $+9/(2C\ln^{2}\!2)$ and curvature $-7/(C\ln^{2}\!2)$, leaving
$-5/(2C\ln^{2}\!2)$ in $\ln F_r$ and $+5/(2\ln^{3}\!2\,C)$ in $\Dnl$. At the
next order,
\begin{eqnarray}
\Dnl&=&\tfrac12\log_2\frac{\pi\ln^{2}\!2\,C}{8}\nonumber\\
&&+\frac{1}{\ln2}\Bigl[\frac{5}{2\ln^{2}\!2\,C}
-\frac{19}{\ln^{4}\!2\,C^{2}}\Bigr]+O(C^{-3}),
\label{eq:series}
\end{eqnarray}
The series is asymptotic. It breaks down once the capacity is no longer large,
and the Gaussian closure takes over there.
 
\subsection{Small capacity: the nearly flat spectrum regime}
 
At small capacity Eq.~\eqref{eq:peak} does not merely lose accuracy, it fails:
the profile is narrower than the octave spacing, the tilted integrals no longer
see a slowly varying $P$, and $P_{\max}$ diverges as the spectrum flattens while
$\Dnl$ goes to zero. The second branch has to come from the discrete spectrum.
Set $d:=2^{\LA}$ and let $p_i=(1+\epsilon_i)/d$ on a dyadic support, with
$\sum_i\epsilon_i=0$ and $\max_i|\epsilon_i|=o(1)$. Only one sector competes:
ranks at most $d/2$ have fidelity at most $(1+o(1))/2$, while the full-rank
branch tends to one and is eventually optimal. Expanding
$\sqrt{p_i}=d^{-1/2}(1+\tfrac12\epsilon_i-\tfrac18\epsilon_i^{2}+\dots)$
gives
\begin{equation}
F_d=\frac1d\Bigl(\sum_i\sqrt{p_i}\Bigr)^{2}
=1-\frac{1}{4d}\sum_i\epsilon_i^{2}+o\Bigl(d^{-1}\sum_i\epsilon_i^{2}\Bigr).
\label{eq:Fflat}
\end{equation}
The first order cancels by $\sum_i\epsilon_i=0$.
 
The same second moment is the capacity. Entanglement energies are
$E_i=\LA-\ln(1+\epsilon_i)/\ln2$, so $E_i-\LA=-\epsilon_i/\ln2+o(\epsilon_i)$
and $C=d^{-1}\sum_i\epsilon_i^{2}/\ln^{2}\!2$ to the same order. It is also a
purity: $d^{-1}\sum_i\epsilon_i^{2}=d\,\tr\rho_A^{2}-1$ is the relative purity
excess, so near flatness $\Dnl$ reads off a purity. Taking the logarithm of
Eq.~\eqref{eq:Fflat},
\begin{equation}
\Dnl=-\log_2F_d\simeq\frac{\ln2}{4}\,C,
\qquad(C\ll1),
\label{eq:nearflat}
\end{equation}
the second line of Eq.~\eqref{eq:caplaw}. Both the trace cancellation and the
exact dyadic support matter: the first kills the linear term, the second makes
the full-rank branch available at all.
The law vanishes on a flat dyadic spectrum, where every $\epsilon_i=0$ and the
block is a collection of Bell pairs. It governs the ergodic steady states of
the main text, whose capacity is $O(1)$.
 
The Haar random state shows both the reach of Eq.~\eqref{eq:nearflat} and its
edge. Set $\Delta:=L-2\LA>0$. Across such an unbalanced cut the reduced state
is nearly maximally mixed, the Born profile collapses onto $\LA$, and the
capacity is $C\to2^{-\Delta}/\ln^{2}\!2$. Equation~\eqref{eq:nearflat} then
returns
\begin{equation}
\Dnl(\infty)=\frac{2^{-\Delta}}{4\ln2},
\label{eq:haar}
\end{equation}
the stationary value quoted in the main text, halving with every unit of
imbalance.
 
The balanced cut is the endpoint of this family and is not analytic. There the
Schmidt spectrum follows the Marchenko--Pastur law, whose capacity is
$C=(\pi^{2}/3-11/4)/\ln^{2}\!2\simeq1.12$ rather than the
$1/\ln^{2}\!2\simeq2.08$ that $\Delta\to0$ in Eq.~\eqref{eq:haar} would suggest.
The profile is hard-edged, $\max_i|\epsilon_i|$ is no longer $o(1)$, and the
half-rank sector takes over from the full-rank branch: Eq.~\eqref{eq:nearflat}
gives $0.195$ against the exact $D_0\simeq0.244$. The hypothesis fails where
the family is broadest, and the Gaussian closure is used there instead.
 
\section{Bounds relating nonlocal magic, the capacity of entanglement, and the
fermionic nonlocal stabilizer entropy}%
\label{app:bounds}
 
We bound $\Dnl$ by the capacity of entanglement and, for fermionic Gaussian states, by the fermionic nonlocal stabilizer entropy $\MF$. The argument runs through the concentration of the Born profile $P(E)$ alone, and no Gaussian closure is assumed at any point.
 
\subsection{Spectral concentration and the capacity bound}
 
The object carrying the argument is the largest Born weight found in a unit window of entanglement energy, $a=\sup_x\Pr_P(x-1<E\le x)$, the windows taken half-open so that the atoms of the discrete spectrum are counted once. For an arbitrary Schmidt spectrum,
\begin{equation}
\frac{a}{8}\le\max_{0\le k\le\LA}F_{2^k}\le8a,
\label{eq:conc}
\end{equation}
so that $\Dnl=-\log_2a+O(1)$. This is the unsmoothed counterpart of the peak-height law, the height of the mode being replaced by the weight of the heaviest unit window; when $P$ is smooth one has $a\simeq P_{\max}$ and the two differ by the constant $\log_2(4/\ln2)$ fixed in Appendix~\ref{app:peak}. Concentration fixes the resource, not the entanglement carried: a translation of $P$ changes the entanglement entropy and leaves $a$ where it was.
 
For the right inequality of Eq.~\eqref{eq:conc} we write the counting function $\mathcal N(x)=\#\{i:E_i\le x\}$. Partitioning $(-\infty,x]$ into unit intervals gives $\mathcal N(x)=\int_{(-\infty,x]}2^E P(E)\,dE\le2^{x+1}a$. Since the Schmidt probabilities are ordered, $\mathcal N(E_i)\ge i$, and therefore $ip_i\le2a$. Inserting this into $F_r=r^{-1}(\sum_{i\le r}\sqrt{p_i})^2$ and using $\sum_{i\le r}i^{-1/2}\le2\sqrt r$, we obtain $F_r\le8a$.
 
The left inequality follows from a single window. Consider $(x-1,x]$ with Born weight $a_x$. Its contribution to the amplitude sum satisfies $\sum_{i\le\mathcal N(x)}\sqrt{p_i}\ge2^{(x-1)/2}a_x$. Rounding $\mathcal N(x)$ up to the nearest dyadic rank costs at most a factor of two, $r\le2\mathcal N(x)$, so that $F_r\ge a_x^2/(8a)$, and letting $a_x$ approach the supremum gives the claim.
 
The capacity enters through a Chebyshev estimate. Since the variance of $P$ is $C$ by definition, at least three quarters of the Born weight lies in a window of width $4\sqrt{1+C}$ centred on $S$, and at most $5\sqrt{1+C}$ unit windows cover that region, so that one of them carries $a\ge3/(20\sqrt{1+C})$. Together with Eq.~\eqref{eq:conc} this gives $\Dnl\le\tfrac12\log_2(1+C)+\alpha_1$ with the explicit constant $\alpha_1=\log_2(160/3)\simeq5.74$. Neither smoothness nor any assumption on the higher cumulants enter this argument.
 
The converse fails. A spectrum carrying a single spike of finite weight beside a broad tail has arbitrarily large capacity and yet $a=O(1)$, hence bounded nonlocal magic. Large capacity forces large nonlocal magic only on an anticoncentrated profile, one in which no window of unit width carries a finite fraction of the weight. Local dynamics produces anticoncentrated bells, the regime assumed throughout.
 
\subsection{Fermionic Gaussian states}
 
We now specialize to fermionic Gaussian states, where a matching bound from below is available. Restricting the local unitaries to Gaussian ones defines $\MF=\min_{U_A^{\rm F},U_B^{\rm F}}\mathcal M_2[(U_A^{\rm F}\otimes U_B^{\rm F})|\Psi\rangle]$~\cite{turkeshi2026local,tirrito2026nonlocal}, with $\mathcal M_2$ the stabilizer R\'enyi entropy~\cite{hamma2022stabilizer,bittel2024stabilizer,piroli2023quantifying}.
 
Bringing the block covariance matrix to its canonical form $\bigoplus_j\nu_j(i\sigma_y)$ we obtain the $\LA$ eigenvalues $\nu_j=|1-2n_j|\in[0,1]$, one per natural orbital, each mode contributing the pair $(1\pm\nu_j)/2$. The entanglement energy is then a sum of $\LA$ independent two-valued terms of gap $\epsilon_j=\log_2[(1+\nu_j)/(1-\nu_j)]$, whose variance is $C=\sum_jc(n_j)$ with $c(n_j)=\tfrac14(1-\nu_j^2)\epsilon_j^2$. The restricted minimization leaves a sum over the same modes~\cite{turkeshi2026local},
\begin{equation}
\MF=\sum_{j=1}^{\LA}f(n_j),\qquad
f(n_j)=-\log_2\bigl(1-\nu_j^2+\nu_j^4\bigr).
\label{eq:mfnl}
\end{equation}
The summand vanishes on Bell modes, $\nu_j=0$, and on product modes, $\nu_j=1$, so that $\MF$ is extensive whenever the mean contribution per mode stays positive. The single-mode capacity being bounded, $C=O(\LA)$, and the capacity bound above already constrains $\Dnl$ to grow at most logarithmically.
 
The lower bound follows from the weighted Kolmogorov--Rogozin inequality~\cite{kesten1969sharper}, which bounds the concentration of such a sum. Applied to the independent contributions with window widths $\min\{1,\epsilon_j/2\}$, the deterministic modes omitted, it gives $a\le{\rm const}\,[1+\sum_j(1-\nu_j)\min\{1,\epsilon_j^2\}]^{-1/2}$, and the single-mode comparison $f(n_j)\asymp(1-\nu_j)\min\{1,\epsilon_j^2\}$ turns this into $a\le{\rm const}/\sqrt{1+\MF}$. With Eq.~\eqref{eq:conc} we obtain $\Dnl\ge\tfrac12\log_2(1+\MF)-\alpha_0$, the constant $\alpha_0$ again universal.
 
A third estimate is immediate: taking per mode the better of the product and the Bell stabilizer, whose fidelity is $q_*(\nu)=\tfrac12[1+\max\{\nu,\sqrt{1-\nu^2}\}]\ge1-\nu^2+\nu^4$, and forming the tensor product of these choices gives $\max_kF_{2^k}\ge\prod_j(1-\nu_j^2+\nu_j^4)=2^{-\MF}$, that is $\Dnl\le\MF$.
 
\subsection{The logarithmic law and its limits}
 
The two logarithms close on each other whenever $C\le K_0\MF$ at fixed $K_0$, in which case
\begin{equation}
\Dnl=\tfrac12\log_2(1+\MF)+O(1),
\label{eq:law}
\end{equation}
with the constant set by $K_0$ and the universal constants above alone. A uniform bound $\epsilon_j\le\bar\epsilon$ on the gaps of the fluctuating modes is sufficient, since the single-mode expressions then give $c(n_j)\le K_0f(n_j)$ uniformly, and no central-limit assumption is needed.
 
The two also become comparable when both carry finite densities, $C=\LA\bar c+o(\LA)$ and $\MF=\LA\bar f+o(\LA)$. The Gaussian quenches of the main text sit in precisely this regime, their occupations relaxing to a smooth profile with a nonvanishing stationary capacity per mode, and Eq.~\eqref{eq:law} then follows without any uniform bound on the individual gaps.
 
The compression of an extensive fermionic magic into its logarithm is, however, not universal, as the following construction shows. Consider $m$ modes with occupations $n_j=1/(r_j+1)$, where $r_j$ is the $j$-th prime, and define $w_m=\prod_{j=1}^m r_j/(r_j+1)$. Each Schmidt probability has the form $p_I=w_m/\prod_{j\in I}r_j$. Unique factorization makes the products in the denominators distinct positive integers. After ordering them, the $i$-th denominator is therefore at least $i$, giving $p_i\le w_m/i$. It follows that $F_r\le4w_m$ for every rank, while $F_1=w_m$, and hence $\Dnl=-\log_2w_m+O(1)$.
 
Expanding the single-mode contribution at large $r_j$ gives
  $f(n_j)=4\log_2(1+1/r_j)+O(r_j^{-2})$.
  Since $\sum_j r_j^{-2}$ converges, the total correction remains bounded
  as $m$ increases. Consequently,
  $\MF=4[-\log_2w_m]+O(1)$ and hence $\Dnl=\MF/4+O(1)$.
  Both quantities diverge as $m\to\infty$: indeed,
  $-\log_2w_m=\sum_{j=1}^m\log_2(1+1/r_j)$,
  each summand behaves as $1/(r_j\ln2)$ for large $r_j$, and the sum
  of the reciprocals of the prime numbers,
  $\sum_{j=1}^m1/r_j$, grows without bound.
No universal constants can therefore make $\Dnl\le A\log_2(1+\MF)+B$ hold for every Gaussian state, although Eq.~\eqref{eq:conc} does force the two measures to diverge together along any family.
 
\section{Nonlocal magic in the kicked Ising model}
\label{app:kim}
 
Does the behavior found in the random circuit survive the replacement of random gates by a fixed Floquet operator? We answer this in the kicked Ising model (KIM)~\cite{prosen1998time,prosen1999dimensional,prosen2002general}, with Floquet operator
\begin{equation}
U_{\rm KIM}=e^{-ib\sum_jX_j}\,
e^{-i\left(h\sum_jZ_j+\sum_jZ_jZ_{j+1}\right)}.
\label{eq:kim}
\end{equation}
We fix the transverse field to $b=(\sqrt5+5)/8$ and the uniform longitudinal field to $h=(\sqrt5+1)/4$, placing the model in a chaotic regime~\cite{huse2013ballistic,sierant2025anticoncentration}. The state evolves as $|\Psi_t\rangle=U_{\rm KIM}^t|\Psi_0\rangle$, with integer time $t$ counting Floquet periods. The evolution operator is fixed, and we average over $200$ random product states $|\Psi_0\rangle=\bigotimes_j(U_j^{(1)}|0\rangle)$, where the $U_j^{(1)}$ are independent Haar-random single-qubit unitaries~\cite{sierant2025anticoncentration}.
 
Figure~\ref{fig:kim}(a) shows that $\Dnl$ grows from zero, reaches a maximum, and relaxes towards the Haar stationary value. The peak time scales linearly with the subsystem size, $\tauM\propto\LA$ [top inset], while the peak height grows logarithmically, $\DM\propto\log_2\LA$ [bottom inset]. Both scales inherited from the capacity therefore survive.
 
In panel (b) we compare the Gaussian surrogate $D_G(S,C)$ against the exact $\Dnl$, the agreement improving with $\LA$ on the unbalanced cuts shown, as the skewness decays. Panel (c) resolves the initial logarithmic growth, $\Dnl\propto\log_2t$, while panel (d) shows exponential relaxation of $\Delta\Dnl=\Dnl-\Dnl(\infty)$, with a rate independent of $\LA$ over the sizes studied. The phenomenology of the random circuit thus extends to chaotic Floquet dynamics, consistent with what has been observed for other  resources~\cite{sierant2025anticoncentration, Aditya2026equivalence}.
 
\begin{figure}[t]
\centering
\includegraphics[width=0.6\textwidth]{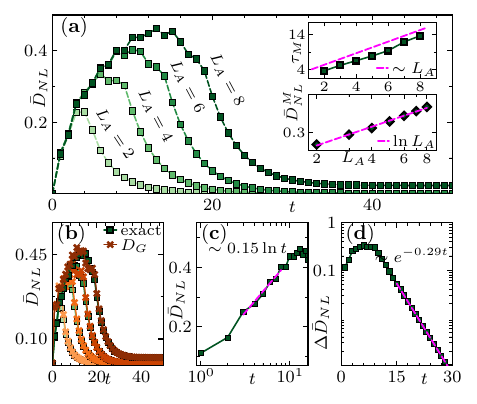}
\caption{\textbf{Nonlocal magic in the kicked Ising model.}
(a)~Rise--peak--fall profile of $\Dnl(t)$ for $L=20$ and $\LA=2,4,6,8$. Insets show the peak-time scaling $\tauM\propto\LA$ (top) and peak-height scaling $\DM\propto\log_2\LA$ (bottom).
(b)~Gaussian estimate $D_G(S,C)$ (diamonds) compared with the exact $\Dnl$ (squares).
(c)~Initial logarithmic growth, $\Dnl\propto\log_2t$.
(d)~Exponential relaxation of $\Delta\Dnl=\Dnl-\Dnl(\infty)$, with a fitted rate of approximately $0.29$ per Floquet period, independent of $\LA$ over the sizes shown.
Results are obtained from exact state-vector simulations averaged over $200$ random initial product states; magenta dashed lines denote fits.}
\label{fig:kim}
\end{figure}
 
\section{ETH prediction for the stationary nonlocal magic of the MFIM}%
\label{app:eth}
In the main text we quoted the stationary nonlocal magic of the MFIM as the
outcome of a high-temperature expansion. Here, we
detail that expansion. The only small parameter is $\beta$: we do not expand in
$\LA/L$, and the condition $\LA\ll L$ enters solely through the applicability of
ETH. We employ natural logarithms in the intermediate steps and convert to bits
at the end, $S$ and $C$ following from their natural-logarithm counterparts on
division by $\ln2$ and $\ln^{2}\!2$.
 
By ETH the block relaxes to $\rho_A(\infty)=e^{-\beta H_A}/Z_A$, with
$Z_A=\mathrm{Tr}\,e^{-\beta H_A}$ and $\beta$ fixed by the energy density of the
initial state. To determine $\beta$, we denote by
$\langle X\rangle_\infty=\mathrm{Tr}\,X/2^{L}$ the infinite-temperature average
on the chain and expand
$\langle H\rangle_\beta=-\partial_\beta\ln\mathrm{Tr}\,e^{-\beta H}$ to first
order, which gives $\langle H\rangle_\beta=-\beta\langle H^{2}\rangle_\infty
+O(\beta^{2})$ since $\langle H\rangle_\infty=0$. Equating this to the mean
energy $E_s$ of the sampled ensemble, we obtain
\begin{equation}
\beta=-\frac{E_s}{\langle H^{2}\rangle_\infty},\qquad
\langle H^{2}\rangle_\infty=(L-1)J^{2}+L(h^{2}+b^{2}),
\label{eq:beta}
\end{equation}
which is negative for the above-median energies of our ensemble.
 
The second equality in Eq.~\eqref{eq:beta} follows from trace orthogonality:
only the square of each term of $H$ is proportional to the identity, the cross
terms between distinct Pauli strings vanishing, so that
$\langle H^{2}\rangle_\infty$ is the sum of the squared couplings weighted by
their multiplicities. The same counting on the block yields the cumulants
$\kappa^\infty_n$ of $H_A$ under $\langle\cdot\rangle_\infty$, which are
cumulants of the Hamiltonian and not of the Born profile $P(E)$. At first order
no term is proportional to the identity, hence $\kappa^\infty_1=0$. At second
order we recover the infinite-temperature variance $V_A$ of the main text, the
multiplicities now being $\LA$ sites and $\LA-1$ bonds, the bond straddling the
cut belonging to neither block. At third order a product of three terms is
proportional to the identity only when a bond term is closed by the two one-site
terms living on its own sites, $(Z_iZ_{i+1})(Z_i)(Z_{i+1})\propto\mathbb{I}$,
which requires the longitudinal field; with the $3!$ orderings, $\kappa^\infty_3$
is proportional to $Jh^{2}(\LA-1)$ and vanishes identically in the free case
$h=0$. One expression, evaluated on $L$ or on $\LA$ sites, fixes both the
temperature and the observable.
 
Since the initial states lie in the middle of the spectrum, $|\beta|\ll1$ and
the cumulant expansion of the partition function converges rapidly,
\begin{eqnarray}
\ln Z_A&=&\LA\ln2+\ln\bigl\langle e^{-\beta H_A}\bigr\rangle_\infty
=\LA\ln2+\sum_{n\ge1}\frac{(-\beta)^{n}}{n!}\,\kappa^\infty_n
=\LA\ln2+\frac{\beta^{2}}{2}V_A-\frac{\beta^{3}}{6}\kappa^\infty_3+O(\beta^{4}).\qquad
\label{eq:lnZ}
\end{eqnarray}
Differentiating Eq.~\eqref{eq:lnZ}, we obtain $\langle
H_A\rangle=-\partial_\beta\ln Z_A=-\beta V_A+\tfrac{\beta^{2}}{2}\kappa^\infty_3$,
which in turn gives the entropy through $S=\ln Z_A+\beta\langle H_A\rangle$. The
higher cumulants of $P(E)$ require no further work. Indeed,
$E_i=-\ln p_i=\beta\varepsilon_i+\ln Z_A$ is an affine function of the
eigenvalues $\varepsilon_i$ of $H_A$, and the Born weights $p_i$ are the thermal
weights themselves, so that every cumulant of $E$ beyond the first is a thermal
cumulant of $H_A$ scaled by $\beta^{n}$, $\kappa_n(E)=\beta^{n}\kappa^{\rm
th}_n(H_A)$. Expanding the thermal cumulants about $\beta=0$ in turn,
$\kappa^{\rm th}_2=V_A-\beta\kappa^\infty_3+O(\beta^{2})$ and $\kappa^{\rm
th}_3=\kappa^\infty_3+O(\beta)$, and converting to bits, we obtain to order
$O(\beta^{4})$:
\begin{eqnarray}
S(\infty)&=&\LA-\frac{1}{\ln2}
\Bigl[\frac{\beta^{2}}{2}V_A-\frac{\beta^{3}}{3}\kappa^\infty_3\Bigr],\nonumber\\
C(\infty)&=&\frac{\beta^{2}V_A-\beta^{3}\kappa^\infty_3}{\ln^{2}\!2},\nonumber\\
\gamma_3(\infty)&=&\frac{\beta^{3}\kappa^\infty_3}{\ln^{3}\!2}.
\label{eq:ethcum}
\end{eqnarray}
These satisfy $S+\tfrac{\ln2}{2}C+\tfrac{\ln^{2}\!2}{6}\gamma_3=\LA$ order by
order, the $\kappa^\infty_3$ contributions combining as
$-\tfrac13+\tfrac12-\tfrac16=0$. The left-hand side is the exponent of the
effective rank $d_G=2^{S+\frac{\ln2}{2}C}$ of the Gaussian closure, so
$d_G=2^{\LA}$ and the full-rank branch wins.
 
We may therefore evaluate $F_r$ at $r=2^{\LA}$ directly. For a
thermal state the amplitude sum is itself a partition function,
$\sum_i\sqrt{p_i}=\sum_ie^{-\beta\varepsilon_i/2}/\sqrt{Z_A(\beta)}
=Z_A(\beta/2)/\sqrt{Z_A(\beta)}$, and no Gaussian closure is required,
\begin{eqnarray}
\Dnl(\infty)=-\log_2\frac{1}{2^{\LA}}\frac{Z_A(\beta/2)^{2}}{Z_A(\beta)}
=\LA-\frac{2\ln Z_A(\beta/2)-\ln Z_A(\beta)}{\ln2} .
\label{eq:ethexact}
\end{eqnarray}
Inserting Eq.~\eqref{eq:lnZ} twice, the extensive parts cancel, the quadratic
terms combine as $\tfrac14-\tfrac12=-\tfrac14$ and the cubic ones as
$-\tfrac{1}{24}+\tfrac16=\tfrac18$, leaving $2\ln Z_A(\beta/2)-\ln Z_A(\beta)
=\LA\ln2-\tfrac{\beta^{2}}{4}V_A+\tfrac{\beta^{3}}{8}\kappa^\infty_3$. Hence,
\begin{equation}
\Dnl(\infty)=\frac{\beta^{2}V_A}{4\ln2}
\Bigl[1-\frac{\beta\,\kappa^\infty_3}{2V_A}\Bigr]+O(\beta^{4}).
\label{eq:ethd}
\end{equation}
The leading term is the stationary value quoted in the main text. It is also
Eq.~\eqref{eq:nearflat} evaluated on Eq.~\eqref{eq:ethcum}, since
$(\ln2/4)\beta^{2}V_A/\ln^{2}\!2=\beta^{2}V_A/(4\ln2)$: the thermal block at
$|\beta|\ll1$ is exactly the nearly flat spectrum of that branch. Moreover, $\beta<0$ while $\kappa^\infty_3>0$, so that the
cubic term adds to the quadratic one rather than cancelling it: the correction
accumulates with $\LA$, and the truncation at $O(\beta^{2})$ underestimates the
stationary value. The entropy is considerably less sensitive to it, being $\LA$
minus a small correction of which $\Dnl$ is itself the leading part, which
explains why the two are affected so differently at the same order in $\beta$.

\section{The Page curve of nonlocal magic for Gaussian states}%
\label{app:page_curve}
Here we derive the Page curve of the nonlocal magic of a random Gaussian state,
which is the stationary state reached by the matchgate circuits of the main
text. The same derivation returns the saturation value of the Gaussian quenches
once the random-matrix occupation density is replaced by the generalized Gibbs
one.
 
To adapt the closed form of $\Dnl$ to product spectra, we note that the reduced state
is $\rho_A=\bigotimes_j\mathrm{diag}(n_j,1-n_j)$, so that the Schmidt values are
the $2^{\LA}$ products $\lambda_{\bm\sigma}=\prod_jq_j^{\sigma_j}$ with
$q^0_j=n_j$ and $q^1_j=1-n_j$, and every R\'enyi partition function factorizes,
$Z_A(\alpha)=\sum_{\bm\sigma}\lambda_{\bm\sigma}^{\alpha}
=\prod_j[n_j^{\alpha}+(1-n_j)^{\alpha}]$. The counting and amplitude functions
of Eq.~\eqref{eq:RA} are then built on this product measure, and the three
regimes below are the three ways of evaluating them.

For a random Gaussian state, the block correlation matrix is the compression of a
  random projector onto $2\LA$ of the $2L$ Majorana coordinates. At fixed
  $\ell=\LA/L$, its eigenvalues have the limiting density~\cite{bianchi2021page}
  \begin{equation}
  \varrho_\ell(n)=\frac{\sqrt{(n_+-n)(n-n_-)}}{2\pi\ell\,n(1-n)},\qquad
  n_\pm=\tfrac12\pm\sqrt{\ell(1-\ell)} .
  \label{eq:wachter}
  \end{equation}
Since the cumulants are sums over modes, they are
extensive at fixed $\ell$, $S=\LA\bar h(\ell)$ and $C=\LA\bar c(\ell)$, with
$\bar h=\int\varrho_\ell\,h$ and $\bar c=\int\varrho_\ell\,c$ the averages of the
single-mode functions of the main text. Both integrals are elementary, the
factor $n(1-n)$ of $c(n)$ cancelling the same factor in the denominator of
Eq.~\eqref{eq:wachter}; specifically
$\bar h(\ell)=[\ln2-1+(\ell-1)\ln(1-\ell)/\ell]/\ln2$, while $\bar c$ is the
stationary capacity per mode of the saturation law, now averaged over
Eq.~\eqref{eq:wachter} rather than over momenta. To characterize the small-cut
regime we record the behavior of $\bar c$ at small $\ell$. Writing $x=2n-1$,
Eq.~\eqref{eq:wachter} becomes a semicircle of radius $2\sqrt{\ell(1-\ell)}$ up
to the factor $1-x^{2}$, whence $\langle x^{2}\rangle=\ell+O(\ell^{2})$; since
$c(n)=x^{2}/\ln^{2}\!2+O(x^{4})$ near $n=\tfrac12$, we obtain
$\bar c\simeq\ell/\ln^{2}\!2$ and hence $C\simeq\LA^{2}/(L\ln^{2}\!2)$.
 
\emph{Small cuts.} For $\LA\lesssim\sqrt L$ the winning sector of the dyadic
maximization is the full-rank one, $k^\star=\LA$, and the prefix is the
whole spectrum. The amplitude sum then factorizes together with the spectrum,
$\sum_{\bm\sigma}\sqrt{\lambda_{\bm\sigma}}=\prod_j(\sqrt{n_j}+\sqrt{1-n_j})$,
so that $F_{2^{\LA}}=\prod_j(\sqrt{n_j}+\sqrt{1-n_j})^{2}/2^{\LA}
=\prod_j[\tfrac12+\sqrt{n_j(1-n_j)}]$ and the nonlocal magic becomes exactly
additive over the modes,
\begin{equation}
\Dnl=\sum_{j=1}^{\LA}d(n_j),\qquad
d(n)=-\log_2\bigl[\tfrac12+\sqrt{n(1-n)}\bigr].
\label{eq:additive}
\end{equation}
This is the only case in which $\Dnl$ is extensive, and $d(n)$ provides an
operational meaning for the nonlocal magic of a single partially occupied
orbital. It vanishes at $n=\tfrac12$, where the mode is a Bell pair and hence a
stabilizer state, and it applies as long as every mode lies inside
$|n-\tfrac12|<1/(2\sqrt2)$, beyond which the rank-one sector overtakes the
full-rank one and the mode is effectively
frozen. 
Expanding $d$ near $n=\tfrac12$ gives $d\simeq x^{2}/(4\ln2)$, so that averaging over  Eq.~\eqref{eq:wachter} at small $\ell$ and using
  $\langle x^{2}\rangle\simeq\ell$ we obtain
  \begin{equation}
  \Dnl\simeq\LA\,\frac{\langle x^{2}\rangle}{4\ln2}\simeq\frac{\LA^{2}}{4L\ln2},
  \label{eq:flank}
  \end{equation}
  used to describe the small-subsystem behavior of matchgate circuits in the
  main text. Comparing with the small-$\ell$
  capacity above, Eq.~\eqref{eq:flank} is Eq.~\eqref{eq:nearflat} evaluated on
  $C\simeq\LA^{2}/(L\ln^{2}\!2)$, as it must be, the small block being nearly
  flat. At fixed $\LA$,
  the nonlocal magic therefore decays as a power of $L$ rather than exponentially,
  since a Gaussian state retains a capacity of order $\LA\ell$ on a small block.
  The full-rank sector keeps winning as long as $F_{2^{\LA}}>F_{2^{\LA-1}}$,
  which for a nearly flat spectrum holds while the additive value of
  Eq.~\eqref{eq:flank} stays below an $O(1)$ threshold, that is while
  $\LA\lesssim\sqrt L$. Beyond that point $k^\star$ drops below $\LA$ and
  Eq.~\eqref{eq:additive} ceases to hold.
 
\emph{Intermediate cuts.} For $\LA\gg1$ the entanglement energy is a sum of many
independent contributions, so that the central limit theorem renders $P(E)$ a
bell of mean $S$ and variance $C$ and the integrals of Eq.~\eqref{eq:RA} become
elementary. Completing the square in the general tilt, we obtain
$\int^{E}2^{\theta E'}P(E')dE'
=2^{\theta S+\frac{\ln2}{2}\theta^{2}C}\,\Phi(z-\theta\sigma)$, with
$z=(E-S)/\sqrt{C}$ and $\sigma=\sqrt{C}\ln2$. Setting $\theta=1$ and
$\theta=\tfrac12$ yields $R$ and $A$, whose ratio is the Gaussian closure quoted
in the main text without proof; the effective rank there is $d_G=R(\infty)=2^{S+\frac{\ln2}{2}C}$,
and $z_r$ follows from $\Phi(z_r-\sigma)=r/d_G$. 
Hence the Page curve is $D_G(\LA\bar h(\ell),\LA\bar c(\ell))$, a dome in $\ell$ whose only inputs are the averages of $h(n)$ and $c(n)$ over the occupation density $\varrho_\ell(n)$. The sole approximation is the neglect of $\gamma_n$ for $n\ge3$, which the additivity over modes controls.

\emph{Large cuts.} Here the capacity is extensive and the series of
Eq.~\eqref{eq:series} applies. At the balanced cut $C=\bar c(\tfrac12)\,L/2$,
so that it reproduces the saturation law of the main text together with its
corrections in $1/L$, the height of the dome growing logarithmically with $L$.
 
At $\ell=\tfrac12$ one
has $n_\pm=1$ and $0$, so that Eq.~\eqref{eq:wachter} degenerates to the arcsine
law $\varrho_{1/2}(n)=[\pi\sqrt{n(1-n)}]^{-1}$. This is also the generalized
Gibbs occupation density of the dimer quench: from
$n_k=(1-\cos k)/2$ with $k$ uniform one has $dk/dn=[n(1-n)]^{-1/2}$, hence
$\varrho(n)=\pi^{-1}dk/dn=[\pi\sqrt{n(1-n)}]^{-1}$. The balanced-cut Page value
of the matchgate circuit and the saturation of the Hamiltonian quenches are
therefore governed by one and the same $\bar c$, as our exact data confirm.
Furthermore, the measured stationary capacity carries a constant offset on top
of $\bar c\,\LA$, a boundary term which Eq.~\eqref{eq:wachter} discards by
construction and which shifts $\Dnl$ only at $O(\LA^{-1})$.
 
\section{Initial states with flat occupations in Gaussian Hamiltonian quenches}%
\label{app:flat_occupations}
The saturation law of the main text requires a nonvanishing stationary capacity
per mode $\bar c$. Here we determine $\bar c$ analytically for quenches from
computational-basis product states, whose generalized Gibbs occupations $n_k$ are
independent of $k$. We call such a profile flat, a property of the mode
occupations that is distinct from the flatness of the entanglement spectrum
responsible for the vanishing of $C$: the two coincide only at half filling,
which is where $\bar c$ vanishes.
 
Consider a number-conserving quadratic quench from a state diagonal in the
computational basis, $C_{jl}(0)=\delta_{jl}m_j$ with $m_j\in\{0,1\}$ and filling
$\nu=L^{-1}\sum_jm_j$. Diagonality of $C(0)$ collapses the double sum in the
momentum occupations,
\begin{equation}
n_k=\langle c^\dagger_kc_k\rangle=\frac1L\sum_{j,l}e^{-ik(j-l)}C_{jl}(0)
=\frac1L\sum_jm_j=\nu ,
\label{eq:flatnk}
\end{equation}
which are therefore independent of $k$. The stationary correlation matrix is the
transform of this constant,
\begin{equation}
C_{jl}(\infty)=\int\frac{dk}{2\pi}\,n_k\,e^{ik(j-l)}=\nu\,\delta_{jl},
\label{eq:flatCjl}
\end{equation}
so that every block orbital carries $n_j=\nu$ and the additivity over modes gives
$S(\infty)=\LA\,h(\nu)$ and $C(\infty)=\LA\,c(\nu)$. Hence $\bar c=c(\nu)$, which is strictly positive for
every $\nu\neq0,\tfrac12,1$. The saturation law therefore holds
unchanged for any such state away from half filling, the filling entering
through the intercept alone; the empty and full cases are unentangled and
trivially carry no nonlocal magic.
 
The single-mode capacity $c(n)=n(1-n)\log_2^{2}[n/(1-n)]$ has a double zero at
$n=\tfrac12$, so that $\bar c$ vanishes at half filling. The N\'eel state
realizes this case. There $\rho_A(\infty)=\mathbb{I}/2^{\LA}$ and $p_i=2^{-\LA}$
for all $2^{\LA}$ Schmidt values, whence
\begin{eqnarray}
F_r&=&\frac1r\Bigl(r\,2^{-\LA/2}\Bigr)^{2}=r\,2^{-\LA},\nonumber\\
\max_{0\le k\le\LA}F_{2^{k}}&=&F_{2^{\LA}}=1,
\label{eq:flatF}
\end{eqnarray}
and $\Dnl(\infty)=0$. The maximally mixed block is the reduced state of $\LA$
Bell pairs, whose spectrum is flat and dyadic, and Eq.~\eqref{eq:nearflat}
returns the same value, every $\epsilon_i$ vanishing and with it $C$.
 
The relaxation rate follows from the residual coherences. For the N\'eel state
the only structure beyond Eq.~\eqref{eq:flatnk} is
$\langle c^\dagger_kc_{k+\pi}\rangle$, which evolves as
$e^{-i(\epsilon_k-\epsilon_{k+\pi})t}$ and is absent from the generalized Gibbs
ensemble. The stationary points of $\epsilon_k-\epsilon_{k+\pi}$ make its
real-space transform decay as $t^{-1/2}$, so that
$\delta n_j=n_j-\tfrac12\sim t^{-1/2}$. Inserting
$c(n)\simeq(2\delta n)^{2}/\ln^{2}\!2$ in the mode sum for $C$,
\begin{equation}
C(t)\propto t^{-1},\qquad
\Dnl(t)\simeq\frac{\ln2}{4}\,C(t)\propto t^{-1},
\label{eq:flatrate}
\end{equation}
in agreement with our numerics.
 
Equation~\eqref{eq:flatnk} requires $C(0)$ to be diagonal, which the dimer state
of the main text is not: its singlets carry coherences between paired sites, so
that $n_k=(1-\cos k)/2$ sweeps the interval $[0,1]$ and $\bar c$ is finite.

\end{widetext}

\bibliographystyle{apsrev4-2}

\bibliography{ref}

\end{document}